\documentclass[prl,superscriptaddress,amsmath,amssymb,twocolumn]{revtex4-2}
\usepackage{natbib}
\usepackage{graphicx}         
\usepackage{dcolumn}          
\usepackage{bm}               
\usepackage{upgreek}
\usepackage{booktabs} 				
\usepackage{tabularx}
\usepackage{array}
\usepackage{ulem}
\usepackage{float}
\usepackage[colorlinks=true,urlcolor=blue,breaklinks=true]{hyperref}
\usepackage{xcolor}
\usepackage[a4paper,left=2cm,right=2cm]{geometry}
\usepackage{orcidlink}
\usepackage{xr}
\usepackage{titlesec}

\newcommand{\alox}{Al$_2$O$_3$ }

\begin{document}
\title{Fabrication and characterization of InAs nanowire-based quantum dot structures utilizing buried bottom gates}

\author{Anton Faustmann}
\affiliation{Peter Gr\"unberg Institut (PGI-9), Forschungszentrum J\"ulich, 52425 J\"ulich, Germany}
\affiliation{JARA-Fundamentals of Future Information Technology, J\"ulich-Aachen Research Alliance, Forschungszentrum J\"ulich and RWTH Aachen University, 52425 J\"ulich, Germany}

\author{Patrick Liebisch}
\affiliation{Institute of Semiconductor Electronics, RWTH Aachen University, Aachen, Germany}

\author{Benjamin Bennemann} 
\affiliation{Peter Gr\"unberg Institut (PGI-9), Forschungszentrum J\"ulich, 52425 J\"ulich, Germany}
\affiliation{JARA-Fundamentals of Future Information Technology, J\"ulich-Aachen Research Alliance, Forschungszentrum J\"ulich and RWTH Aachen University, 52425 J\"ulich, Germany}

\author{Pujitha Perla} 
\affiliation{Peter Gr\"unberg Institut (PGI-9), Forschungszentrum J\"ulich, 52425 J\"ulich, Germany}
\affiliation{JARA-Fundamentals of Future Information Technology, J\"ulich-Aachen Research Alliance, Forschungszentrum J\"ulich and RWTH Aachen University, 52425 J\"ulich, Germany}

\author{Mihail Ion Lepsa}
\affiliation{Peter Gr\"unberg Institut (PGI-9), Forschungszentrum J\"ulich, 52425 J\"ulich, Germany}
\affiliation{JARA-Fundamentals of Future Information Technology, J\"ulich-Aachen Research Alliance, Forschungszentrum J\"ulich and RWTH Aachen University, 52425 J\"ulich, Germany}

\author{Alexander Pawlis}
\affiliation{Peter Gr\"unberg Institut (PGI-9), Forschungszentrum J\"ulich, 52425 J\"ulich, Germany}
\affiliation{JARA-Fundamentals of Future Information Technology, J\"ulich-Aachen Research Alliance, Forschungszentrum J\"ulich and RWTH Aachen University, 52425 J\"ulich, Germany}

\author{Detlev Gr\"utzmacher}
\affiliation{Peter Gr\"unberg Institut (PGI-9), Forschungszentrum J\"ulich, 52425 J\"ulich, Germany}
\affiliation{JARA-Fundamentals of Future Information Technology, J\"ulich-Aachen Research Alliance, Forschungszentrum J\"ulich and RWTH Aachen University, 52425 J\"ulich, Germany}

\author{Joachim Knoch}
\affiliation{Institute of Semiconductor Electronics, RWTH Aachen University, Aachen, Germany}

\author{Thomas Sch\"apers}
\email{th.schaepers@fz-juelich.de}
\affiliation{Peter Gr\"unberg Institut (PGI-9), Forschungszentrum J\"ulich, 52425 J\"ulich, Germany}
\affiliation{JARA-Fundamentals of Future Information Technology, J\"ulich-Aachen Research Alliance, Forschungszentrum J\"ulich and RWTH Aachen University, 52425 J\"ulich, Germany}
\hyphenation{}
\date{\today}

\begin{abstract}
Semiconductor nanowires can be utilized to create quantum dot qubits. The formation of quantum dots is typically achieved by means of bottom gates created by a lift-off process. As an alternative, we fabricated flat buried bottom gate structures by filling etched trenches in a Si substrate with sputtered TiN, followed by mechanical polishing. This method achieved gate line pitches as small as 60 nm. The gate fingers have low gate leakage. As a proof of principle, we fabricated quantum dot devices using InAs nanowires placed on the gate fingers. These devices exhibit single electron tunneling and Coulomb blockade.
\end{abstract}
\maketitle

\section{Introduction}

In recent years, various approaches to create host systems for qubits have made considerable progress in terms of gate fidelity, coherence times, and scalability \cite{Chatterjee21}. In this context, a striking and particularly elegant way to create a quantum mechanical two-level system that forms the quantum bit is to use an electron with spin 1/2 confined in a quantum dot \cite{Burkard23}. Based on this, significant efforts have been made to build more complex systems comprising coupled quantum dot qubits. The direct exchange interaction resulting from particle exchange, achieved by reducing the potential barrier, is typically limited to directly neighboring qubits. A particular challenge for the implementation of a flexible quantum circuit design is to achieve coherent coupling of quantum dots over a longer distance. This can be gained, for example, by using floating gate electrodes \cite{TriDia12} or by spatially reallocating a qubit itself \cite{Xue24}. Additionally, two mechanisms using a superconducting electrode connected to both qubits have also been proposed, i.e. by crossed Andreev reflection \cite{LeiFle13} or superconductor-mediated exchange interaction \cite{HasCat15,Rosado21}.

In semiconductor systems, electron confinement to form a quantum dot can be achieved by heterostructures of materials with different band gaps, such as GaAs/AlGaAs or Si/SiGe, forming a two-dimensional electron system in combination with gate electrodes \cite{Burkard23}. Alternatively, quantum dots can also be formed by using stronger confined semiconductor structures, i.e. different types of nanowires, which greatly simplifies the required gating \cite{FasFuh07,NadFro10,PetMcF12,BerNad13,Hee19,MuHua21,ZhaWu22}. In this approach, the nanowire is placed on a set of densely packed parallel gate fingers to define quantum dots along the nanowire. Their pitch should be in the range of $60-100$\,nm \cite{FasFuh05,FasFuh07,ZhaWu22}. To form these bottom gates, the metal electrodes are usually deposited in a lift-off process on a substrate passivated by a dielectric layer. To isolate the nanowire above while maintaining sufficient capacitive coupling, the gates are typically covered with a high-k dielectric layer. Due to the fabrication process, the resulting surface is not completely flat. Apart from the inevitable valleys between the gate lines, there is a certain roughness of the metal itself, which is likely to lead to defects and charge traps in the covering dielectric. This is highly undesirable since gate performance is severely degraded when the dieletric is prone to charge rearrangements \cite{Hee19,WeiWir14,RibMit05}. 

As an alternative approach to gate structures fabricated by metal lift-off, we prepared buried bottom gate structures. Here, the actual gates consist of trenches etched into the Si substrate and filled with sputtered TiN superconductor. A subsequent mechanical polishing step is used to achieve a smooth surface of the gates and the surrounding silicon. After deposition of a dielectric layer, these totally flat gate structures can be used as a platform to define quantum dot structures in a semiconductor nanowire placed on top. Due to the smooth and flat surface, we expect improved gate performance. Here, we describe in detail the fabrication and characterization of buried gate structures. Subsequently, nanowire-based QD structures are fabricated and measured, where the QD potential is defined by buried gate fingers below the nanowire. 

\section{Experimental}

\subsection{Fabrication of Polished Bottom Gates}

For the buried bottom gate process, the gate layout is first etched into a high resistivity Si(100) substrate using a SF$_6$ reactive ion etching (RIE) process to a depth of 120\,nm. Electron beam lithography using a 100\,nm thick CSAR (Product name by manufacturer Allresist) resist allows for a gate line pitch of 60\,nm. In order to provide sufficient isolation between the individual gate lines, the substrates are dry oxidized to achieve a SiO$_2$ layer of about 20\,nm thickness covering the trenches and the plane surface. The gates are filled with sputtered TiN with a nominal thickness of 140\,nm. Due to columnar growth with increased growth rates near the edges of the trenches, the nominal film thickness is chosen to be greater than the depth of the trenches. A mechanical polishing step using a SiO$_2$ abrasive is applied to achieve a flat surface. Polishing must be stopped when the excess TiN between the gate lines has been removed. At the same time, enough material must be left in the trenches. (Details on the polishing step are given in Supplementary Material.) Figure~\ref{FAB:BGP_NachPolieren} shows a gate structure after the polishing step. The structure of the gate lines and the Si substrate form a flat surface. The different materials and the insulating SiO$_2$ in between can be resolved. The process allows to choose contact materials that are difficult to be used for a lift-off process, such as tungsten and generally sputter-deposited metals.
\begin{figure}
\includegraphics[width=1.00\columnwidth]{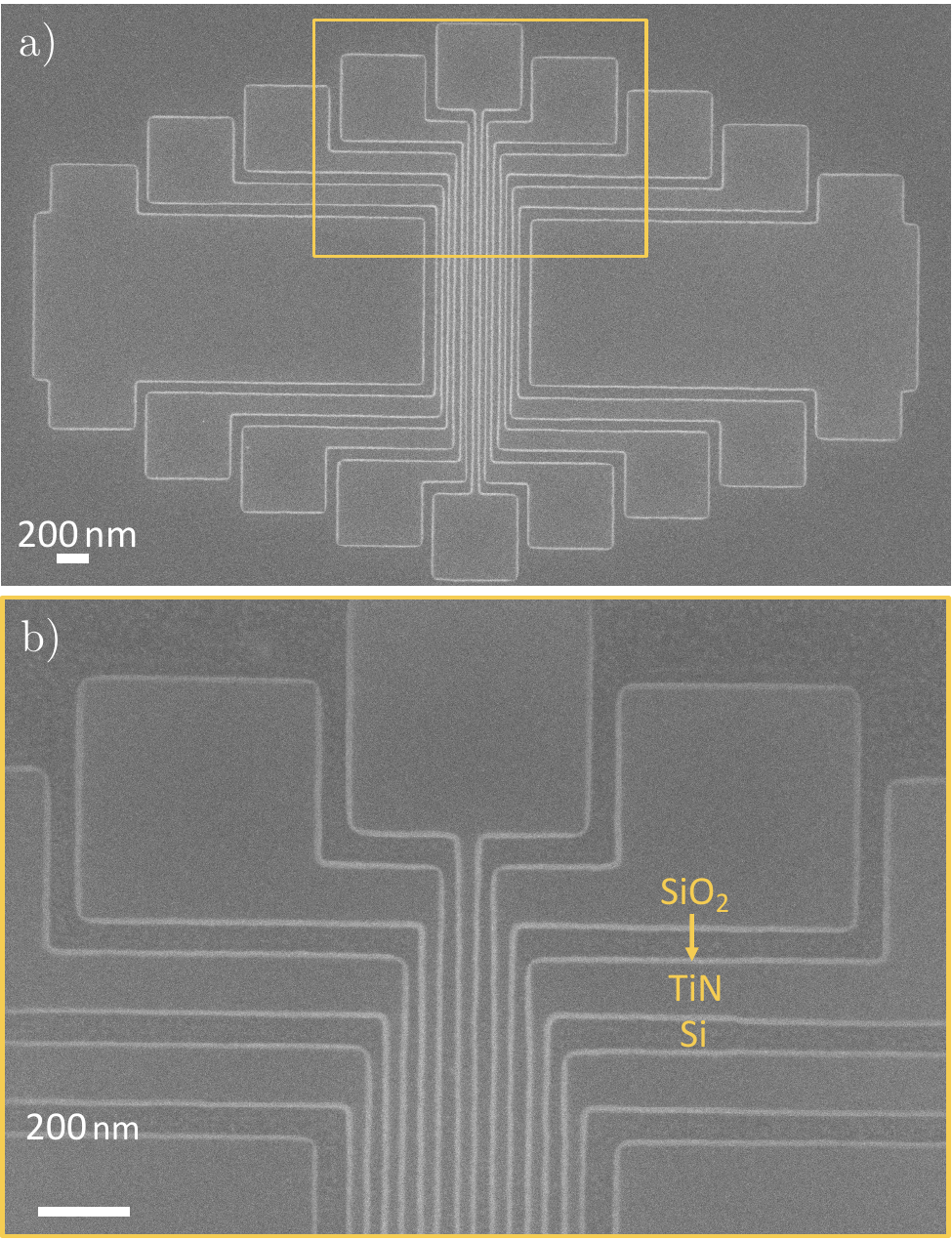}
\caption{a) Scanning electron micrograph of a buried bottom gate structure after polishing. Gate stripes are visible in the center. Gates and surrounding silicon form a flat surface. b) Close-up of a). TiN gates, the Si substrate, and SiO$_2$ insulation interlayers can be clearly distinguished. }
\label{FAB:BGP_NachPolieren}
\end{figure}

\begin{figure}
\includegraphics[width=1.0\columnwidth]{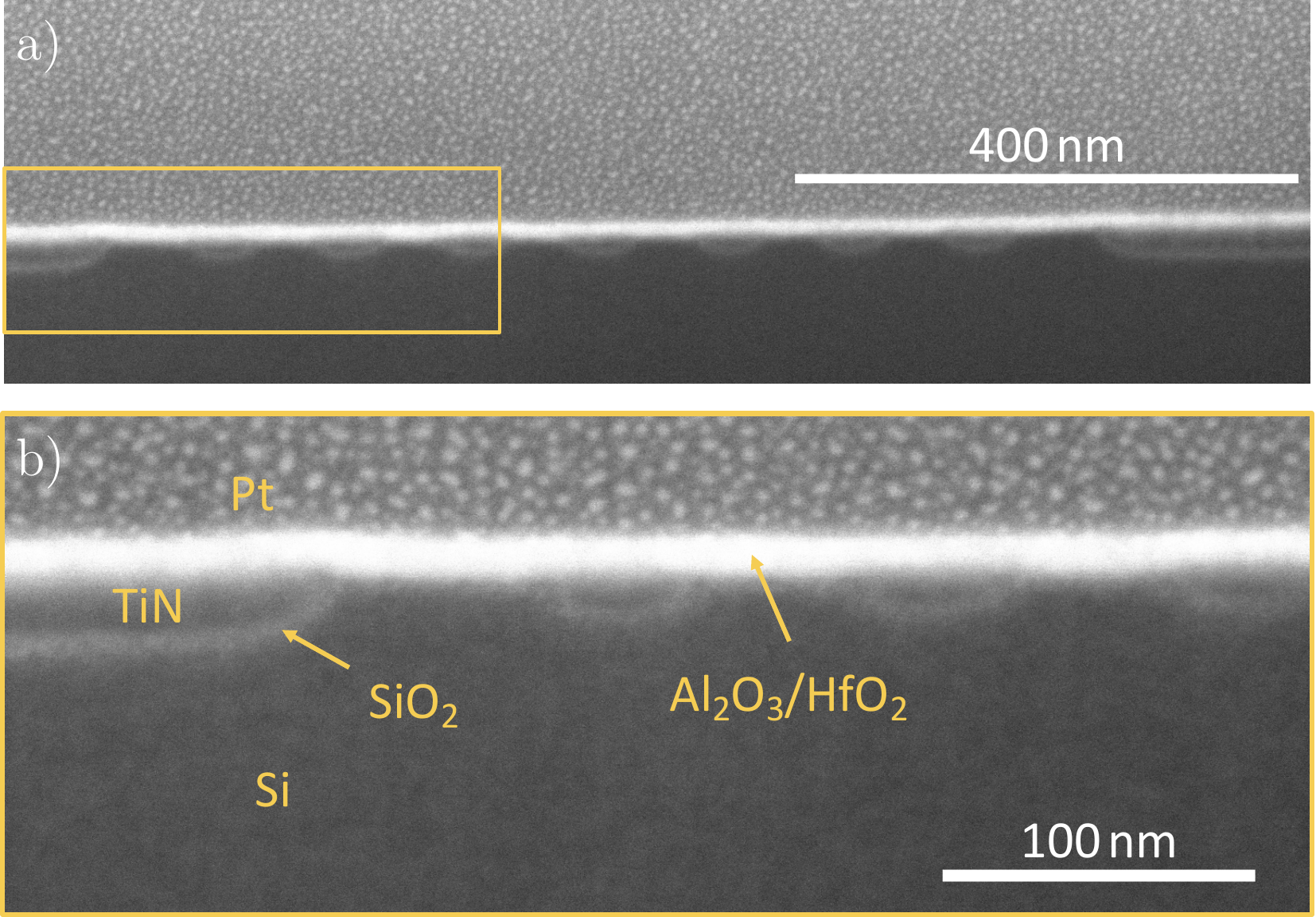}
\caption{a) and b) Scanning electron micrograph of FIB cut through polished buried bottom gates. The Pt layer was deposited during the FIB process for experimental reasons.}
\label{FIB_BGP}
\end{figure}

After the polishing step, the gate structures are covered with a dielectric layer composed of 3\,nm  \alox and 13\,nm HfO$_2$, deposited by atomic layer deposition (ALD). A cross-sectional view on a focused ion beam (FIB) cut through gate lines is shown in Fig.~\ref{FIB_BGP}. The image was taken by a scanning electron mircoscope (SEM). The depicted device was fabricated with a gate pitch of 100\,nm. The rounded shape of the gate lines and the greater depth of the larger pad compared with the gates are caused by the RIE etching process (see Supplementary Material).
Due to the presence of the covering dielectric layer, forming electric contacts to the gates requires the opening of contact windows. This is performed by RIE using a process based on CHF$_3$ (see Supplementary Material for details). 

\subsection{Quantum Dot Device Fabrication}

To evaluate the performance of buried gate fingers, semiconductor nanowire-based quantum dot structures are fabricated. In our case, molecular beam epitaxy (MBE) grown InAs nanowires with a diameter of about 120\,nm are used. The nanowires were in-situ covered with a 20-nm-thick superconducting aluminum half-shell separated by a full-shell 2.5-nm-thick \alox barrier layer. The nanowire structure is motivated by the coupling of two spin-qubits by a superconductor at a later stage \cite{LeiFle13,HasCat15}.

The nanowires are placed on top of the bottom gate structures with previously opened contact windows and are aligned perpendicular to the gate lines. Piezo-based micromanipulation inside a SEM was used for this purpose. In this way, it is possible to place the nanowires with the Al half-shell oriented upwards, thus avoiding shielding of the gate potentials by the Al half-shell (details on the nanowire transfer are provided in the Supplementary Material). A second lithography step is used to define the source and drain contact regions at both ends of the nanowire, where the Al half-shell and the underlying \alox layer are partially removed in order to contact the semiconductor core. In our case with aluminum as the superconducting material, this is done by a wet chemical process using a TMAH-based lithography developer \footnote{MF-CD 26} and CSAR resist. 

\begin{figure}
\centering
\includegraphics[width=1.0\columnwidth]{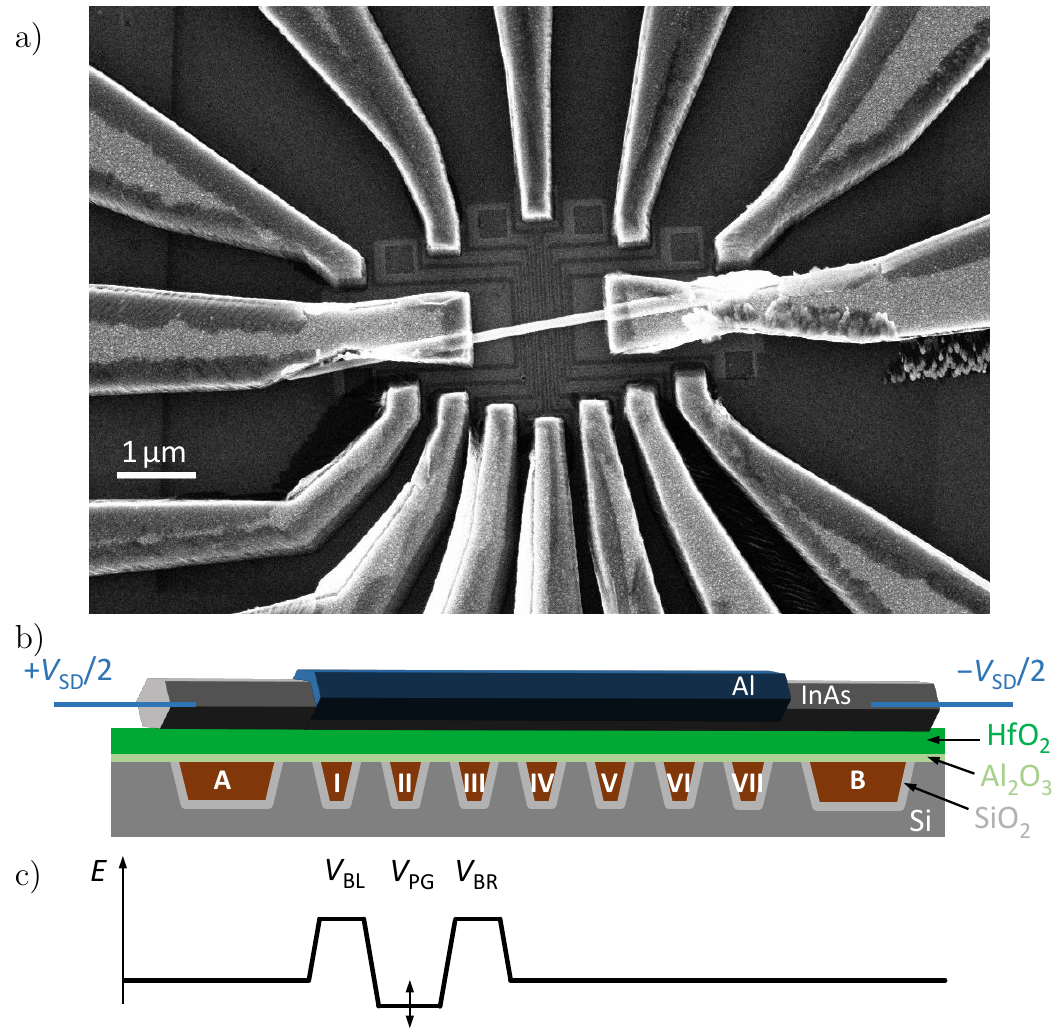}
\caption{a) Scanning electron micrograph of device for creating quantum dots. b) Schematic of gate layout and electric wiring to investigate Coulomb blockade in the nanowire. A triplet of bottom gates (I-III) is used for defining two tunnel barriers ($V_\mathrm{BL}$ and $V_\mathrm{BR}$) and one intermediate plunger gate $V_\mathrm{PG}$ to tune the potential in the quantum dot. The source-drain voltage at the nanowire ends is applied symmetrically around common ground. c) Schematic of the potential landscape along the nanowire axis. The arrows symbolise the tuning of the plunger gate in order control the number of electrons in the dot.}
\label{FIG:SchemaQD}
\end{figure}

As a final step, contacts to the ends of the InAs nanowire and to the bottom gates are fabricated in a metal lift-off process using a Ti/Pt (130\,nm/90\,nm thickness) layer stack with in-situ Ar sputtering to enhance contact quality. Contacting the gates and the InAs nanowire in a single step has proven advantageous to avoid damage from electrostatic discharges during the process. A scanning electron microscope image and a schematic cross section of the final device structure are shown in Figs.~\ref{FIG:SchemaQD} a) and b), respectively. The final design layout features a nanowire core that is ohmically contacted at both ends and tunnel-coupled to its Al half-shell by an \alox layer. In addition, the nanowire is separated by the gate dielectric, which covers the seven lower gate fingers along its axis. Each gate is individually contacted, allowing the desired potential landscape to be set for quantum dot definition in the nanowire (see Fig.~\ref{FIG:SchemaQD}~c). We investigated three devices, i.e. device A, B, and C. 

The electrical measurements on the characteristics of the gate electrodes and on the quantum dot devices were carried out in a $^3$He cryostat at a temperature $T \approx 600\,$mK and a dilution refrigerator at $T \approx 60\,$mK.

\section{Electrical Characterisation}

\subsection{Leakage current measurements}

\begin{figure}
\hspace{0.05\columnwidth}a)\hspace{0.44\columnwidth}b)\hspace{0.43\columnwidth}\,
\centering
\includegraphics[width=1.0\columnwidth]{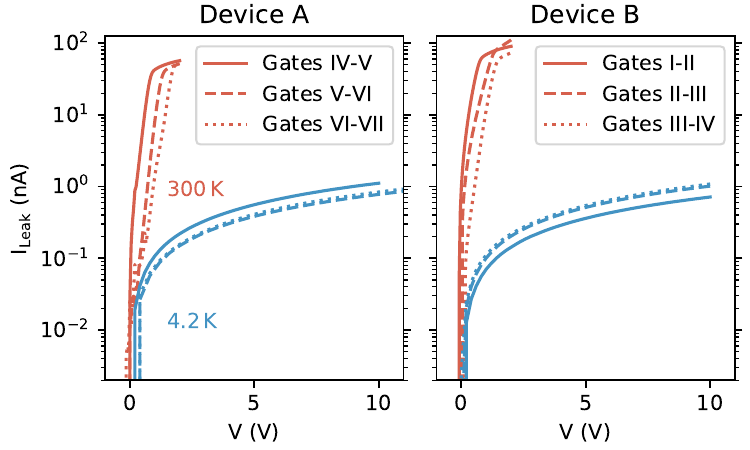}
\caption{a) and b) Leakage currents of two buried bottom gate substrates between neighbouring gate stripes for devices A and B, respectively. At room temperature, considerable leakage currents occur above approx. $0.5\,$V bias, reaching the order of 100\,nA at 2\,V. At $4.2\,$K, maximum leakage reaches $1-2\,$nA at voltages of $\pm 10\,$V.
}
\label{EXP:BGP16_Leakage}
\end{figure}

Electrical measurements were performed to assess the quality of the insulation between the individual gate fingers on the fabricated substrate. At room temperature, buried gates were found to have proper isolation between nearest neighbours up to about $\pm 0.5\,$V (see Fig.~\ref{EXP:BGP16_Leakage}). At higher voltages of about 2\,V, leakage currents in the order of 100\,nA were observed. At cryogenic temperatures, however, leakage was found to be greatly reduced. Bias voltages of $\pm 10\,$V resulted in leakage currents of about 1\,nA at 4.2\,K. No destruction of gates or the covering dielectric due to excess currents was observed within this voltage range. Since the quantum dot devices are intended to operate at temperatures below 1\,K, the leakage between the gates is not considered detrimental to their application.
\subsection{Formation of Quantum Dots in Nanowires}

As a next step, we demonstrate that the set of buried gates is suitable to form quantum dots in the InAs nanowire. The source-drain bias $V_\mathrm{SD}$ was applied symmetrically around common ground in order to avoid distortion. The DC voltage bias was overlaid with a small AC oscillation and analysed by a lock-in amplifier in order to determine differential conductance in response to both $V_\mathrm{SD}$ and plunger gate voltage V$_\mathrm{PG}$. Triplets of bottom gates were used to define a quantum dot in the nanowire, with the intermediate one serving as the plunger gate and the two outer ones negatively biased to form tunnel barriers. All other gates were biased positively to provide good conductance in the nanowire and its contacts. For this reason and due to the tunnelling-dominated transport, the contribution from the quantum dot is expected to be largely dominant, suggesting the validity of the measurements in two-terminal configuration. The corresponding potential profile is shown schematically in Fig.~\ref{FIG:SchemaQD} c). By making use of the plunger gate, the electrostatic potential in the dot can be altered with respect to the environment. By appropriate choice of the potentials, the system can be tuned to provide precisely one energetically accessible electron state in the dot \cite{Kou01,HanKou07,ZwaDzu13}. Both effective voltages and currents at the gate electrodes were continuously monitored in order to exclude leaking situations.

\begin{figure}
\hspace{0.07\columnwidth}a)\hspace{0.6\columnwidth}b)\hspace{0.2\columnwidth}\,
\centering
\includegraphics[width=1.0\columnwidth]{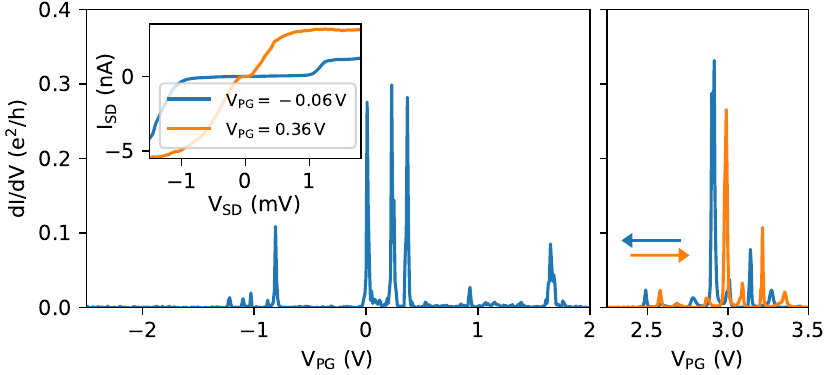}
\caption{a) Differential conductance through a nanowire quantum dot (device C, gates V-VII) depending on the plunger gate voltage at fixed source-drain bias of $100\,$µV at $60\,$mK. The voltage at the barrier gates was kept at $-4.5$\,V. The inset shows corresponding $IV$ curves at fixed gate potentials for $V_\mathrm{PG}=-0.06$\,V and $0.36\,$V, respectively. Sharp increases of conductance at specific gate voltages are observable, with suppressed transport in between. b) Similar measurement (device C, gates III-V) at $V_\mathrm{SD}=-200$\,µV, above induced superconducting gap. The system shows gate hysteresis with the conductance features shifted depending on the direction and speed of the plunger gate alteration.}
\label{FIG:QG_G_VG}
\end{figure}

Figure~\ref{FIG:QG_G_VG}~a) shows the differential conductance through a quantum dot in device C at 60\,mK, plotted over the static plunger gate potential at a constant symmetric source-drain bias of $100\,$µV. For this measurement, the quantum dot was formed between gates V and VII used for creating the barriers with gate VI acting as plunger gate.
At specific values of $V_\mathrm{PG}$, sharp peaks in the conductance are observed, whereas transport is largely blocked in between, which can be attributed to single-electron tunnelling and Coulomb blockade in the quantum dot, respectively. The inset in Fig.~\ref{FIG:QG_G_VG}~a) shows the corresponding current-voltage ($IV$) traces at two different plunger gate voltages, with voltage plateaus in the Coulomb blockade regime. In Fig.~\ref{FIG:QG_G_VG} b), the conductance behaviour of another quantum dot (device C, gates III-V, biased at $V_\mathrm{SD}=200\,$µV) is shown for both directions of the plunger gate voltage variation. The features are shifted by an offset of about $0.08\,$V in gate voltage, giving rise to an hysteresis effect of the gate electrode. This is most likely caused by trapped charges at interfaces and in the HfO$_2$ dielectric \cite{WeiWir14,RibMit05}.

\begin{figure}
\centering
\includegraphics[width=1\columnwidth]{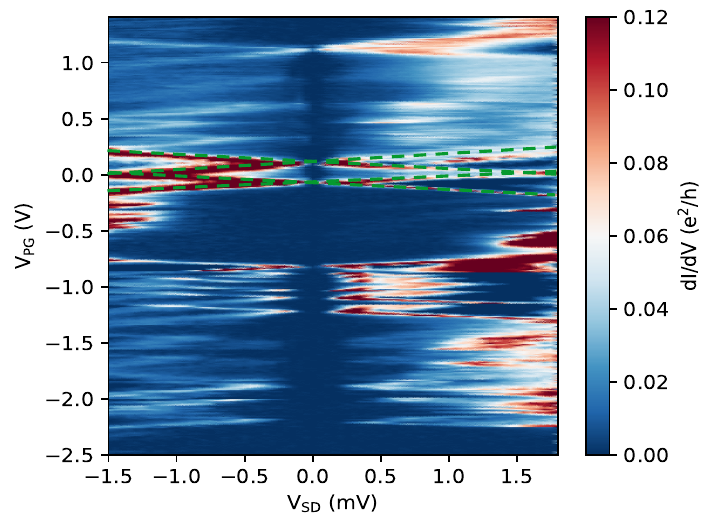}
\caption{Stability diagram for device C measured at $T \approx 60\,$mK. For this measurement, gate VI was used as plunger gate with the tunnel barriers formed by gates V and VII kept at $-4.5\,$V. Features of single electron tunnelling, manifested in typical Coulomb diamond structures, are clearly visible. One example is highlighted by dashed green lines. The measurement has been corrected for sudden jumps in the effective plunger gate potential.}
\label{FIG:CoulombDiamondsBGP16_A2_corrJumps}
\end{figure}

A stability diagram taken at a temperature of 60\,mK showing differential conductance of device C with respect to both plunger gate voltage and source-drain bias is depicted in Fig.~\ref{FIG:CoulombDiamondsBGP16_A2_corrJumps}. The quantum dot was formed with gate lines V and VII for defining the tunnel barriers at $-4.5\,$V and gate VI as plunger gate (cf. Fig.~\ref{FIG:SchemaQD}). Characteristic Coulomb diamonds of blocked transport can clearly be observed. The diagram has been corrected for sudden jumps in the plunger gate potential by removing double features.
The induced superconducting gap in the semiconductor due to the aluminium shell is manifests in an increase of the blocked voltage bias window by $2\Delta/e$ and therefore to avoided closing of Coulomb diamonds at the charge degeneracy points \cite{DohFra08,GroJoer09,FraKou10}. The strength of this effect can only be estimated from the present data and is further complicated by transport remainig possible close to charge degeneracy points due to Andreev reflection \cite{GroJoer09}. An upper bound of the induced gap can be estimated to be around $2\Delta \lesssim 0.15$\,meV, which is considerably below the bulk Al gap, reflecting the weak coupling including the barrier layer. The electrostatic situation of the nanowire quantum dot is complex since contributions from at least five gate electrodes, the effect of the dielectric and gate hysteresis need to be taken into account. Assuming the capacitive charging energy to dominate over quantum confinement energies in the given dimensions, the capacitive energy associated with adding an additional electron into the dots is at about 1.5\,meV, which is in the expected range for comparable geometries of gate-defined quantum dots in InAs nanowires \cite{Hee19}. The plunger gate lever arm $\alpha=-C_\mathrm{PG}/C_\mathrm{tot}$, the ratio between the plunger gate capacitance and the total capacitance is extracted from the slope of the Coulomb diamonds $V_\mathrm{SD}/\Delta V_\mathrm{PG}$, to about 0.009.

The quantum dot characteristics of device C was also determined for various triplets of gate lines defining the quantum dot. The corresponding stability diagrams can be found in the Supplementary Material. The basic parameters such as charging energy $E_\mathrm{C}$ and gate lever arms $\alpha$ for the different devices and gate triplets are summarized in Table~\ref{RES:WerteQD}. The charging energies for all investigated quantum dots observed are in the same area around $1.5-2\,$meV. However, the plunger gate lever arm shows strong deviation between the individual gates, ranging from about 0.01 to 0.05 at different positions along one nanowire. This suggests that the detailed electrostatic situation for each dot varies considerably. The gate lever arms are lower than observed for nanowires using LaLu$_2$O$_3$-isolated top-gates \cite{Hee19}. In general, surface states in InAs can both lead to formation of parasitic quantum dots and influence the effective gating potential \cite{Hee19, WeiWir14}.
\begin{table}
    \centering
    \begin{tabular}{c|c|c|c|c|c}
         Device & Quantum dot & $T$ (mK) & $V_\mathrm{PG} (V)$ &  $E_\mathrm C$ (meV) & $\alpha$\\
         \hline
         C &V-VII & 60 &$\approx 0$  & 1.5 & 0.009\\
         \hline
         C &V-VII & 60 & $\approx -1.4$  & 2.1 & 0.008\\
         \hline
         C &III-V & 60 & $\approx 2.35$  & 1.8 & 0.045\\
         \hline
         C &III-V & 60 & $\approx 2.31$  & 1.65 & 0.04\\
         \hline
         C &III-V & 60 & $\approx 2.25$  & 1.6 & 0.05\\
         \hline
         D &I-III & 600 & $\approx -7.45$  & 2.0 & 0.025\\
         \hline
    \end{tabular}
    \caption{Overview of extracted values for some of the investigated quantum dots: Device, gate triple defining the quantum dot, temperature $T$, plunger gate voltage $V_\mathrm{PG}$ where the dot parameters are taken, extracted charging energy $E_\mathrm{C}$, and lever arm $\alpha$ of plunger gate.}
    \label{RES:WerteQD}
\end{table}

\section{Conclusion}

Buried bottom gate strutures were fabricated and used to form quantum dots in InAs nanowires. By overfilling RIE etched Si trenches with sputtered superconductor TiN and mechanical polishing, smooth gate lines of the same level as the substrate with pitches down to 60\,nm and in between leakage of less than 1\,nA at $\pm 10$\,V at low temperature were achieved. Quantum dot devices based on InAs nanowires placed on the gate fingers exhibited single electron tunneling and Coulomb blockade. A small gate hysteresis was observed, indicating a minor influence of charged traps in the gate dielectric. In our study, we have clearly demonstrated the suitability of buried gate structures for quantum dot structures based on semiconductor nanowires.

The shown measurements were obtained from devices with non-ideal polishing quality and remaining roughness of the gates. Using optimal quality gates structures is likely to provide further enhanced performance. As a next step, it would be interesting to investigate the coupling of quantum dots by means of a tunnel-coupled superconducting electrode, as proposed by Leijnse and Flensberg \cite{LeiFle13} and by Hassler \textit{et al.} \cite{HasCat15}. In addition, the buried gate platform holds the potential to reduce the dimensions of gates and spacings to values below those presented here. In a modification of the process, starting from a doped substrate allows to gain an additional gate section at half pitch by employing the intermediate spaces between the gate fingers. The smoother surface compared with lifted metal gates might be particularly suitable for gating two-dimensional materials such as graphene or transition metal dichalcogenites, see e.g. Ref.~\cite{Mueller2016}.

\section{Acknowledgment}

We thank Herbert Kertz and Christoph Krause for technical assistance. Dr. Florian Lentz and Dr. Stefan Trellenkamp are gratefully acknowledged for their help with the e-beam lithography. The focused ion beam cut has been performed by E. Neumann. The sample fabrication has been performed in the Helmholtz Nano Facility at Forschungszentrum J\"ulich \cite{Albrecht2017}. This work was partly funded by Deutsche Forschungsgemeinschaft (DFG, German Research Foundation) contract numbers KN545/22 and SCHA 835/9-1. 




%

\clearpage
\widetext

\titleformat{\section}[hang]{\bfseries}{\MakeUppercase{Supplemental Note} \thesection:\ }{0pt}{}

\setcounter{section}{0}
\setcounter{equation}{0}
\setcounter{figure}{0}
\setcounter{table}{0}
\setcounter{page}{1}
\renewcommand{\thesection}{\arabic{section}}
\renewcommand{\thesubsection}{\Alph{subsection}}
\renewcommand{\theequation}{S\arabic{equation}}
\renewcommand{\thefigure}{\arabic{figure}}
\renewcommand{\figurename}{Supplemental Figure}
\renewcommand{\tablename}{Supplemental Table}
\renewcommand{\bibnumfmt}[1]{[S#1]}
\renewcommand{\citenumfont}[1]{S#1}

\begin{center}
\textbf{\large Supplementary Material: Fabrication and characterization of InAs nanowire-based quantum dot structures utilizing buried bottom gates}
\end{center}

\section{Polishing results}

In Supplementary Figure~\ref{FIG_BGP_PolishQuality}, gate structures filled with TiN are depicted after the polishing step. Due to the uneven material abrasion, the different gate structures on one single wafer have varying degrees of polishing. Typically, the polishing rate is highest at the center of a wafer. In Supplementary Figure~\ref{FIG_BGP_PolishQuality}~a), an underpolished structure is depicted. Here, the polishing did not remove sufficient material, such that the trenches within the sputtered TiN gate lines are still present. Therefore, in this case the surface exhibits remaining roughness. The opposite situation is shown in Supplementary Figure \ref{FIG_BGP_PolishQuality}~b). For this structure, the abrasion continued until the TiN in the gate lines was almost entirely removed and the SiO$_2$ below became visible. In Supplementary Figures~\ref{FIG_BGP_PolishQuality}~c) and d), two gate structures covered with Al$_2$O$_3$ and HfO$_2$ are shown. The underpolished gates in Supplementary Figure~\ref{FIG_BGP_PolishQuality}~c) exhibit remaining roughness that is expected to compromise their functionality. An ideally polished structure is depicted in Supplementary Figure~\ref{FIG_BGP_PolishQuality}~d). Here, gates and surrounding silicon form an approximately flat surface.
\begin{figure}[ht!]
\centering
\includegraphics[width=0.9\columnwidth]{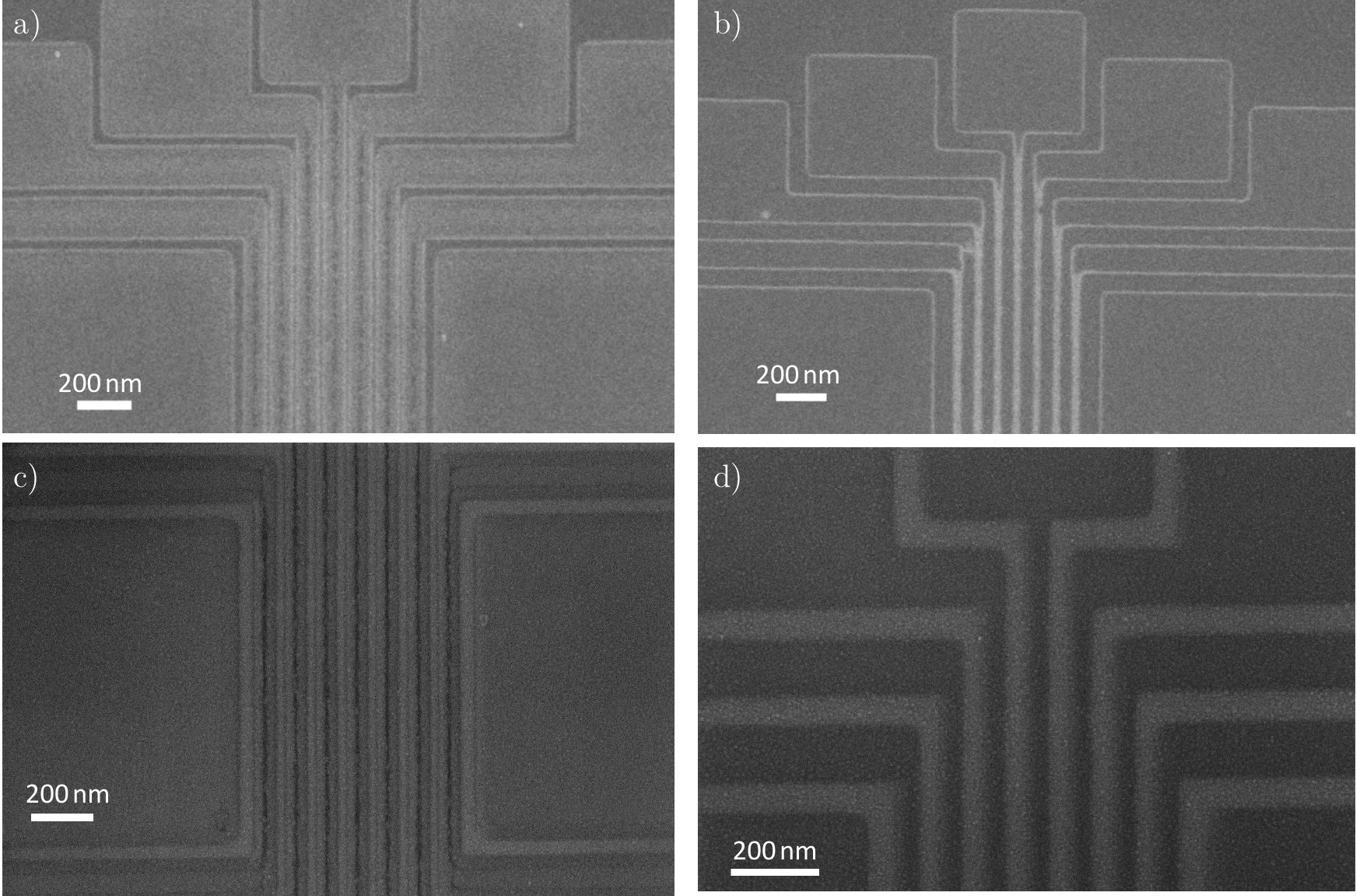}
\caption{a) Underpolished bottom gate structure. The intitial topography from the non-uniform TiN sputtering overfilling the trenches were not fully canceled out leaving back valeys in the center of the stripes visible as darker spots.
b) In contrast to a), too much material has been removed. The insulating SiO$_2$ at the bottom of the gate fingers is largely uncovered and the gates interrupted. c) Close-up of underpolished gates covered with Al$_2$O$_3$/HfO$_2$ dielectric. The remaining roughness is clearly observable as valleys above the gate stripes. d) Close-up of an ideally polished gate structure covered with the dielectric layer. The resulting surface is flat.}
\label{FIG_BGP_PolishQuality}
\end{figure}

\section{Preparation of contact windows and leads}

\begin{figure}
\centering
\includegraphics[width=1.0\columnwidth]{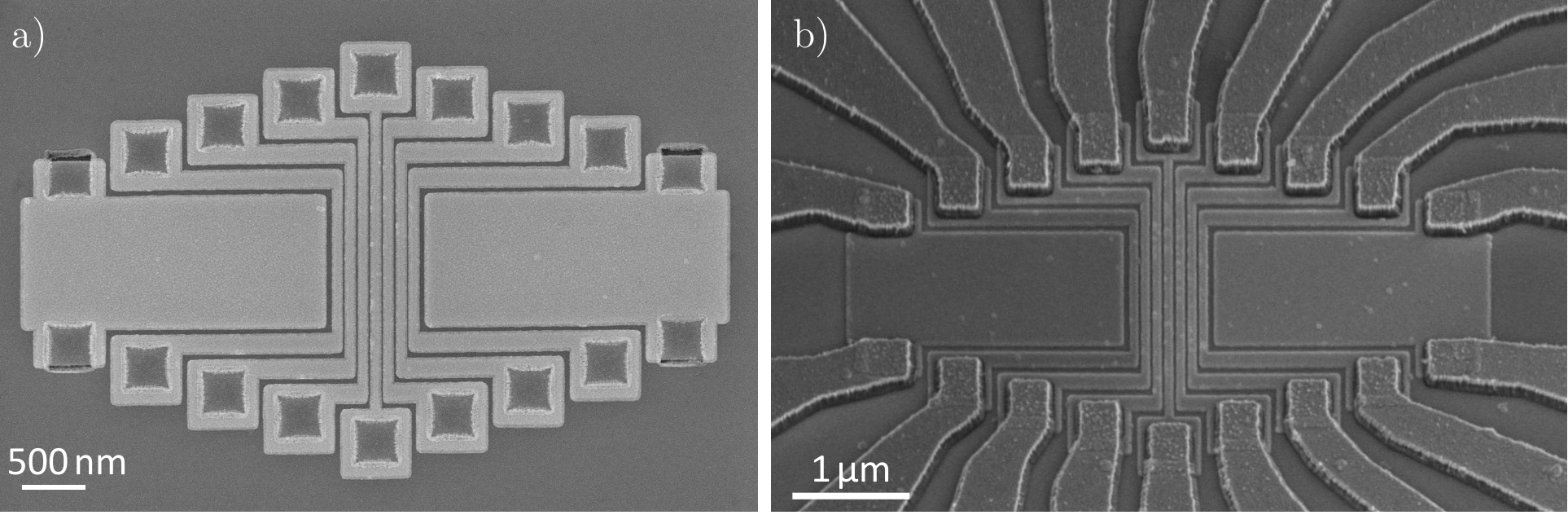}
\caption{a) Bottom gates (deposited by lift-off) with contact windows etched into the dielectric. b) Bottom gates after Ti/Pt leads deposition. Contacting from both sides was only applied to confirm proper electric connecting.
}
\label{FAB:BG_Kontaktfenster}
\end{figure}

Supplementary Figure~\ref{FAB:BG_Kontaktfenster}~a) shows a gate structure with etched windows in the dielectric on the contact pads using a reactive ion etching (RIE) process using a combinantion of CHF$_3$ and O$_2$ plasma. For this sample the bottom gates were defined by a lift-off process. The etching duration was tuned to match the thickness of the oxide layer. The contact fingers are formed in a standard electron beam lithography metal lift-off process from a Ti/Pt layer stack (130\,nm/90\,nm thickness) process using CSAR resist and AR 600-546 \footnote{product names by manufacturer Allresist} developer. An O$_2$ plasma etching step and in-situ Ar-sputtering before metal deposition are essential to ensure good contact quality. A structure contacted with metallic leads is depicted in Supplementary Figure ~\ref{FAB:BG_Kontaktfenster} b). For measurement devices, contacting of both nanowires and gates was done in a single process step since this proved beneficial in order to avoid ESD damage.

\section{Nanowire transfer by micromanipulator}

\begin{figure}[ht]
\includegraphics[width=1.0\columnwidth]{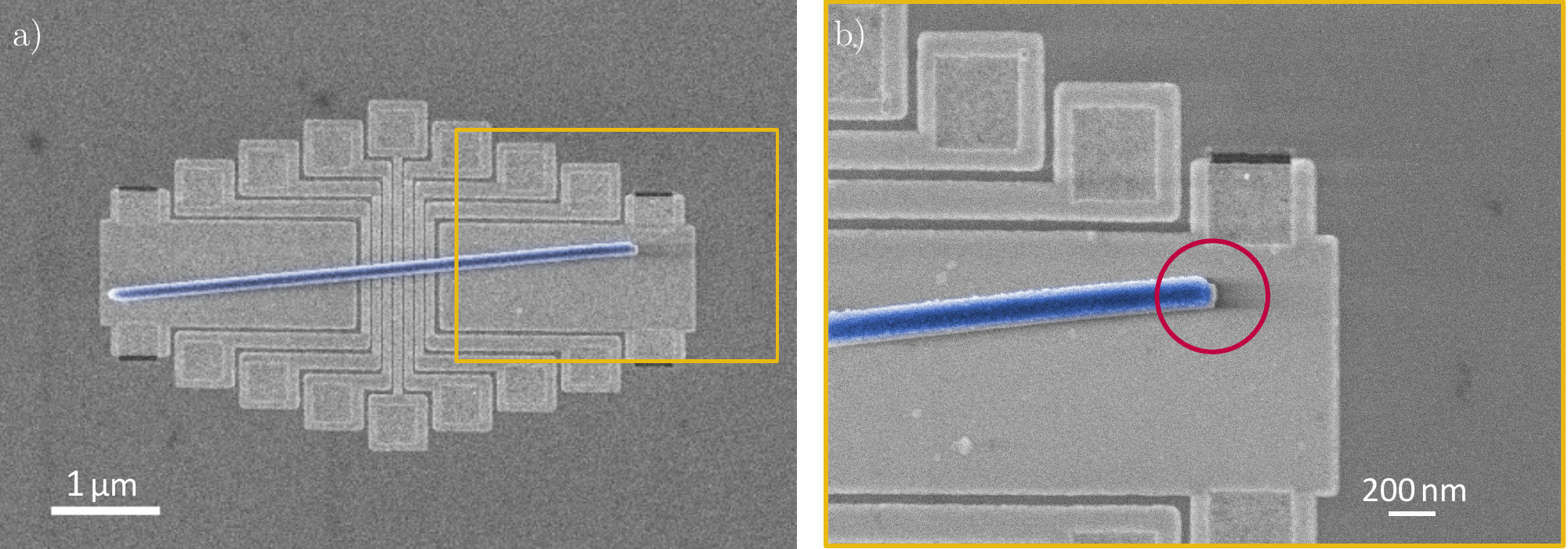}
\caption{a) Nanowire transferred onto a substrate with gate structures prepared by metal evaporation and lift-off. A micromanipulator mounted inside a SEM is used for the transfer. Care must be taken to orient the nanowire with its half-shell upside, as confirmed by the seed point of the wire in b) (marked by the red circle, shell highlighted in blue for clarity). This prevents electrostatic shielding of the gate potential by the aluminum shell.}
\label{FAB:Manip}
\end{figure}

\begin{figure}[ht]
\includegraphics[width=1.0\columnwidth]{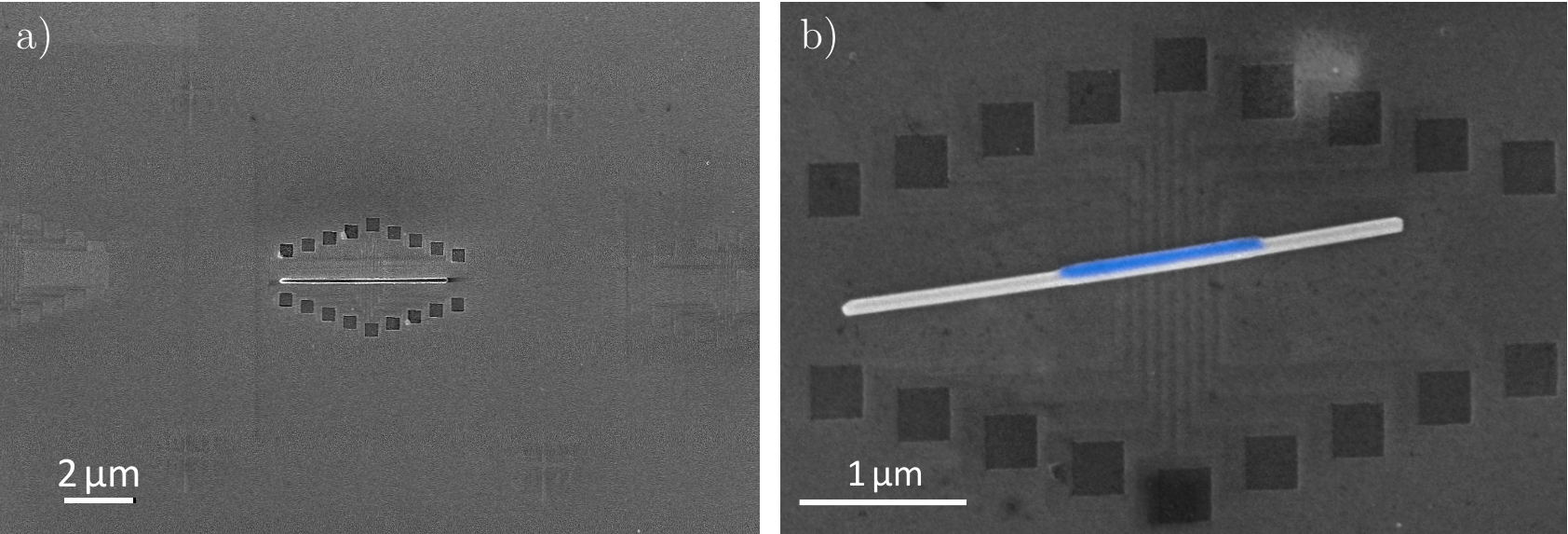}
\caption{a) Scanning electron microscope image of a nanowire transferred to a substrate with buried bottom gates and contact windows opened beforehand. b) Close-up of the gate area. The nanowire placed on the structure already has its aluminum shell partially removed at its ends for the deposition of contact leads to its core. Due to the flat surface, the contrast between the gates and the surrounding silicon is low.}
\label{FAB:BGP_Substrat}
\end{figure}

A piezo-based micromanipulation inside a scanning electron microscope (SEM) was used to place the nanowires on top of the bottom gate structures with previously opened contact windows. This technique had already been used in the past to selectively transfer template-grown nanowires \cite{PerFon21}. Supplementary Figures~\ref{FAB:Manip} and \ref{FAB:BGP_Substrat} show examples of transferred nanowires on substrate with gate fingers prepared by metal deposition and lift off as well as for buried gates, respectively. The nanowires are oriented perpendicular to the gate lines. In the Supplementary Figure \ref{FAB:Manip} it can be seen that the nanowires are placed with the Al half-shell oriented upwards, thus avoiding shielding of the gate potentials by the Al half-shell.

\newpage 
\section{Additional Quantum Dot Measurements}

\begin{figure}
\centering
\includegraphics[width=0.6\columnwidth]{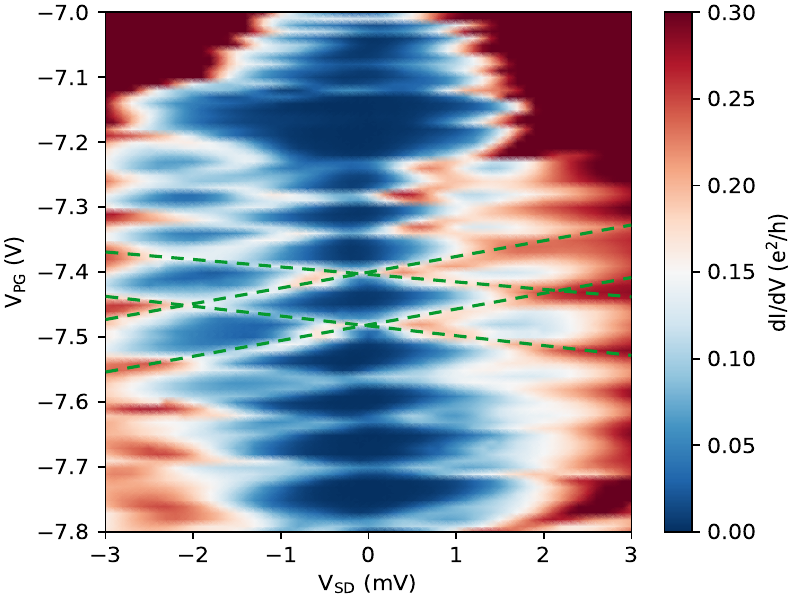}
\caption{Differential conductance $dI/dV $ in units of $G_0 = e^2/h$ through the nanowire with electrostatically defined quantum dot, depending on source-drain bias $V_\mathrm{SD}$ and plunger gate potential $V_\mathrm{PG}$ at $T=600\,$mK. Suppressed transport in a gate-dependent window around $V_\mathrm{SD} = 0\,$V shows Coulomb diamonds. Each diamond corresponds to a specific number of electrons in the dot, increasing by one towards more positive plunger gate voltages. The marked diamond shows a charging energy $E_\mathrm C$ of approx. 2.0\,meV. The extracted gate lever arm $\alpha$ is 0.025.}
\label{FIG:CoulombDiamonds}
\end{figure}
Supplementary Figure~\ref{FIG:CoulombDiamonds} shows a stability diagram of device D taken at 600\,mK. The quantum dot was formed with gate lines I and III for defining the tunnel barriers at $-5\,$V and gate II as plunger gate. Assuming the capacitive charging energy to dominate over quantum confinement energies in the given dimensions, the capacitive energy associated with adding an additional electron into the dots is at about 2.0\,meV. The plunger gate lever arm $\alpha=-C_\mathrm{PG}/C_\mathrm{tot}$, the ratio between the plunger gate capacitance and the total capacitance is extracted from the slope of the Coulomb diamonds $V_\mathrm{SD}/\Delta V_\mathrm{PG}$, to about 0.025.
In this case, the nominal voltage of the plunger gate below $-7\,$V is lower than the barriers at $-5\,$V. This could suggest that the quantum dot observed here is actually not formed as intended by barrier gate electrodes. 

\begin{figure}
\centering
\includegraphics[width=0.6\columnwidth]{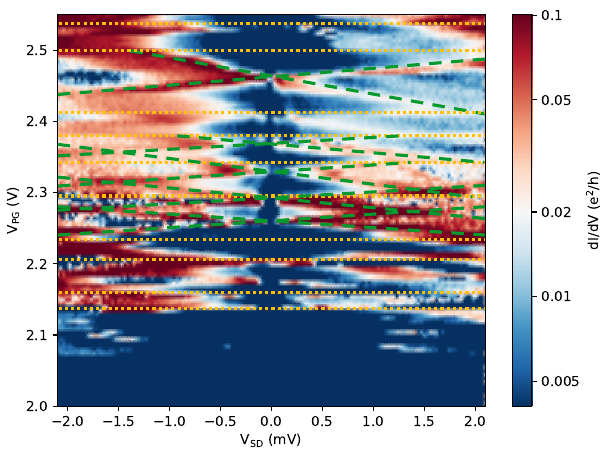}
\caption{Stability diagram of the same device as in main text Figure~6 at 60\,mK, but with gate IV used as plunger gate and III and V for creating the tunnel barriers. 
The latter were biased at $-5\,$V with unused gates at $2$\,V. A number of Coulomb diamonds can be observed with charging energies around $1.7-1.9$\,mV and gate lever arm $\alpha$ of $0.038 - 0.045$. This deviates notably from $\alpha \approx 0.01$ for gate VI. Below a plunger gate voltage of about 2.14\,V transport is strongly reduced and entirely blocked below 2.07\,V.}
\label{FIG:CoulombDiamondsBGP16_A2b}
\end{figure}

Supplementary Figure~\ref{FIG:CoulombDiamondsBGP16_A2b} shows a stability diagram from the same device as in main text Figure~6, but with using gates III-V for defining the quantum dot. The Coulomb diamonds observed shows charging energies of about $1.7-1.9$\,meV and a gate lever arm in the range of 0.038 to 0.045.

%


\end{document}